\documentclass[aps,prl,showpacs,twocolumn,superscriptaddress]{revtex4-1}

\usepackage{graphicx}
\usepackage{color}\usepackage{enumerate}
\usepackage{amsmath}\usepackage{amssymb}\usepackage{stmaryrd}
\usepackage{multirow}
\usepackage{float}
\usepackage{longtable,booktabs}
\usepackage{hyperref}\usepackage{url}
\usepackage{soul} 
\def\beq{\begin{equation}}
\def\eeq{\end{equation}}

\begin{document}

\title{Finding matrix product state representations of highly-excited eigenstates of many-body localized Hamiltonians}

\author{Xiongjie Yu}
\affiliation{Department of Physics, University of Illinois at Urbana-Champaign, IL 61801, USA} 

\author{David Pekker}
\affiliation{Department of Physics and Astronomy, University of Pittsburgh, PA 15260, USA}

\author{Bryan K.  Clark}
\affiliation{Department of Physics, University of Illinois at Urbana-Champaign, IL 61801, USA} 

\begin{abstract}
A key property of many-body localization, the localization of quantum particles in systems with both quenched disorder and interactions,  is the area law entanglement of even highly excited eigenstates of many-body localized Hamiltonians. Matrix Product States (MPS) can be used to efficiently represent low entanglement (area law) wave functions in one dimension. An important application of MPS is the widely used Density Matrix Renormalization Group (DMRG) algorithm for finding ground states of one dimensional Hamiltonians. Here, we describe two algorithms, the Shift and Invert MPS (SIMPS) and excited state DMRG which finds highly-excited eigenstates of many-body localized Hamiltonians. Excited state DMRG uses a modified sweeping procedure to identify eigenstates whereas SIMPS is a shift-inverse procedure that applies the inverse of the shifted Hamiltonian to a MPS multiple times to project out the targeted eigenstate.  To demonstrate the power of these methods we verify the breakdown of the Eigenstate Thermalization Hypothesis (ETH) in the many-body localized phase of the random field Heisenberg model, show the saturation of entanglement in the MBL phase and generate local excitations.

\end{abstract}
 
\maketitle
Many-body localization (MBL) is a dynamical phase transition that happens at finite energy density for an isolated quantum system with interactions and quenched disorder \cite{Basko2006,Basko2007,Oganesyan2007,Pal2010,Aleiner2010,Monthus2010,Imbrie2014}, see also review~\cite{Nandkishore2015}. Many-body localized phases are believed to have a number of properties including: (a) zero conductivity at finite temperature, (b) Poisson statistics of their many-body eigenvalues, (c) eigenstates that obey the area law, (d) an extensive number of local integrals of motion in cases where there isn't a mobility edge, and (e) eigenstates that fail to obey the eigenstate thermalization hypothesis (ETH) and hence MBL systems which fail to thermalize~\cite{Bauer2013,Serbyn2013,Huse2014}. While the ground state wave-function is the key quantity for identifying equilibrium quantum phases, the finite-energy density eigenstates are the analogous identifying feature for dynamical phases.  Analytical arguments for the existence of MBL phases and phase transitions rely primarily on analyzing the highly excited eigenstates using either diagrammatic re-summation~\cite{Basko2006,Basko2007,Oganesyan2007,Pal2010,Aleiner2010,Ros2015} or the real space renormalization group~\cite{Vosk2013,Vosk2014,Pekker2014,Vosk2015,Potter2015}. From the numerical perspective, getting interior eigenvalues (and corresponding eigenstates) using exact diagonalization requires both an exponentially large amount of computer time and memory, and as a result these are limited to rather small (at most 22-site \cite{Luitz2015}) systems~\cite{Pal2010,Berkelbach2010,Iyer2013,Kjall2014}. Time evolution DMRG studies, on the other hand, are limited by the logarithmic-in-time growth of entanglement entropy that occurs in the localized phase~\cite{Znidaric2008,Bardarson2012,Serbyn2013a}.  It has been recently shown that the entire spectrum of eigenstates of a MBL system can be efficiently described by a matrix product  operator (MPO)~\cite{Chandran2014,Pekker2014a}; the best available algorithm for optimizing this MPO~\cite{Pollmann2015} captures features of the entire spectrum but doesn't describe individual states to high fidelity. 
While the bulk of MBL research has been theoretical, there has been considerable recent progress on the experimental side~\cite{DErrico2014,Meldgin2015}. For the first time, clear signs of the MBL transition were observed in an ultracold atom experiment ~\cite{Schreiber2015}. 

In this work, we take advantage of the fact that eigenstates of many-body localized systems obey the area law and can therefore be efficiently represented as a Matrix Product State (MPS)~\cite{Bauer2013,Swingle2013,Friesdorf2014} to develop a set of numerical algorithm for generating a MPS representations of these eigenstates. We use the eigenstates constructed using these algorithms to test the basic properties of MBL, the break-down of ETH, the saturation of the entanglement entropy, and the existence of a large number of local excitations, in the regime of large one-dimensional systems that were previously inaccessible due to the limitations of exact diagonalization. 

The algorithms we develop fall into two broad classes: In class (1), we modify the DMRG sweeping procedure~\cite{Schollwock2011} to pick one of the excited eigenstate of the effective Hamiltonian at each step and hence arrive at an excited eigenstate of the full Hamiltonian. We call this algorithm Excited State-DMRG (ES-DMRG). In class (2), we target a specific energy $\lambda$ by repeatedly applying the operator $(H-\lambda)^{-1}$ to the state vector. We call this algorithm shift-and-invert MPS (SIMPS). SIMPS is enabled by the \textit{inverse-DMRG} algorithm, that we develop here, for efficiently computing the single application of the propagator $(H-\lambda)^{-1}$ to a MPS.

Throughout this manuscript, we use the following one-dimensional Hamiltonian
\begin{align}
H = \sum_i \vec{S}_i \cdot \vec{S}_{i+1} + \sum_i h_i S_i^z
\label{eq:H}
\end{align}
where $h_i$ are sampled uniformly from $[-W,W]$, which is known to have the entire many-body spectrum become localized for $W\gtrsim3.5$ \cite{Pal2010,Luitz2015}. 
After developing our algorithms and showing that they generate eigenstates, we use the eigenstates to test a number of physical properties.
First, we verify that in the many-body localized phase the eigenstate thermalization hypothesis (ETH) breaks down. We explicitly demonstrate this using eigenstates in a tight energy window that we find using SIMPS for $L=30$ chains. 
The second key feature that we verify is the saturation of the entanglement entropy with system size. We verify that this saturation is occurring using eigenstates obtained via SIMPS and ES-DMRG for $L=30,40$.
The third key feature of MBL phases, that we investigate are the local excitations. We show how, given an eigenstate, we can construct approximate eigenstates that differ from a reference eigenstate locally. While this procedure does not have enough power to construct a full set of mutually orthogonal local integrals of motion, it does produce a large number of of ``pretty good'' eigenstates that can potentially be used to investigate the local integrals of motion.

\textbf{Overcoming the exponential gap:}
As a physical system approaches the thermodynamic limit, the inter-level spacing decreases exponentially. Therefore exponentially many states exist within any fixed energy window $(E-\delta,E+\delta)$. The fundamental difficulty in capturing eigenstates, then, is resolving a single eigenstate from a large superposition of eigenstates taken from this window. 


\begin{figure}[th]
\centering
\includegraphics[width=0.85\columnwidth]{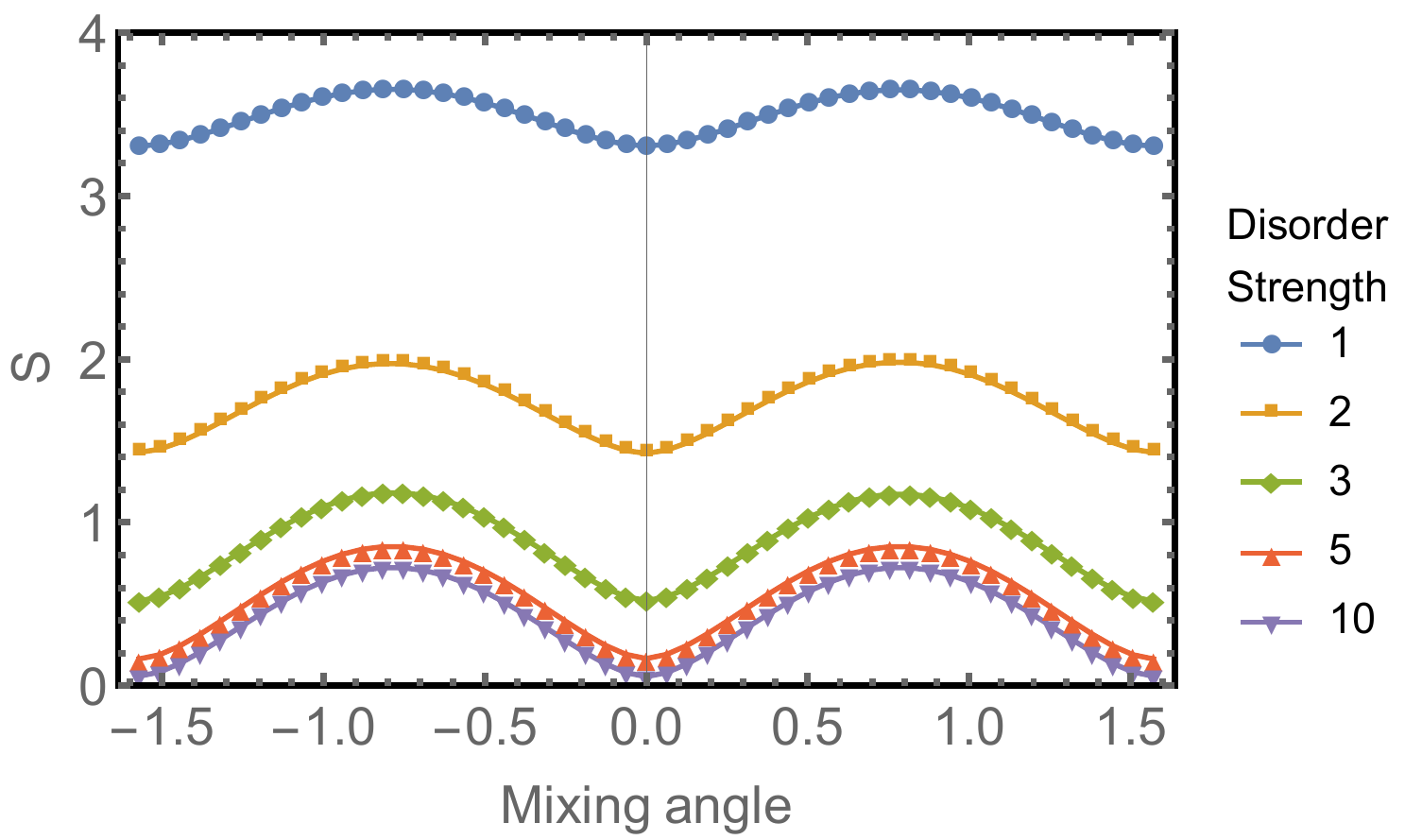}
\caption{Average mid-bond entanglement entropy of the wave-function $\Psi_i= \cos(\alpha) \psi_i + \sin(\alpha) \psi_{i+1}$ as a function of the mixing angle $\alpha$, where $\psi_i$ and $\psi_{i+1}$ are consecutive eigenstates taken from the middle of the spectrum of the Hamiltonian in Eq.~\eqref{eq:H}. Notice that pure eigenstates are local minima of the mean entanglement entropy even when approaching no disorder. The average is over 1\% of eigenstates on a 14 site chain; a single realization of disorder was used for each disorder strength.}
\label{fig:wiggles}
\end{figure}

While it is well known that MBL eigenstates (MBLE) obey area laws~\cite{Bauer2013}, we show that MBLE are local minima in entanglement with respect to linear superpositions of eigenstates nearby in energy (see Fig.~\ref{fig:wiggles}). Conceptually this happens because nearby MBLE are typically localized in different parts of the Hilbert space and building a superposition of them produces a cat state with larger entanglement. This feature  gives an additional metric from which we can identify the MBLE states from the nearly degenerate subspace:  we want to select the minimally entangled states or equivalently MPS with low bond-dimension in this space. This realization allows us to develop successful numerical algorithms for studying these eigenstates. 

\textbf{Excited state DMRG:} The first of these algorithms, while conceptually interesting, is simple to describe and has the advantage of requiring minimal modifications to a current DMRG code. A standard DMRG code sweeps over sites; when working on site $i$, an effective Hamiltonian $H_{i}$ is generated which involves tracing over the other auxiliary and physical degrees of freedom. This effective Hamiltonian $H_{i}$ is then diagonalized and the parameters on site $i$ are replaced with the ground state of $H_{i}$. In our new approach, instead of considering only the ground state of $H_{i}$ we consider all its eigenvalues (of which there are $D^{2}$) replacing the parameters of site $i$ with the one of these eigenvalues. In particular,  we select the eigenvalue with energy closest to the energy of the current MPS to minimize the amount of change in the state as our algorithm progresses. We then follow the typical approach of sweeping back and forth through all the sites. 
As in normal DMRG, we find that a proper starting configuration and bond-dimension protocol can enhance the efficiency of the algorithm keeping it from being stuck and allowing it to more widely sample excited states.  In particular, we typically start with the algorithm in a product state in the $S_z$ basis of bond-dimension two and slowly ramp up the bond-dimension increasing it by one every few sweeps. This ensures we find low bond-dimension states at the energy at which the algorithm converges.  From these sweeps, we take the state we find with lowest variance.  This algorithm scales as typical DMRG.  While currently we are exactly diagonalizing the effective Hamiltonian, a standard shift-and-invert procedure would allow the eigenstate of the effective Hamiltonian to be found without a full diagonalization.  While this approach is powerful, it has some undesirable properties.  In particular, there is no clean way to target a particular energy. 

\textbf{Shift and Invert MPS (SIMPS):}
Our goal is to target a particular energy $\lambda$. The simplest approach is to use standard DMRG techniques to find the ground state of $(H-\lambda)^2$ instead of the ground state of $H$. Unfortunately this decreases an already exponentially small gap in nearby eigenvalues by squaring it.  As  techniques such as ITEBD require propagating to imaginary time larger then this inverse gap, they are unfeasible; we have found sweeping methods are able to converge to some eigenstate but appear to find one further from the energy target than the SIMPS approach described below.

Exact diagonalization is also plagued by the small gap coming from working with $(H-E)^2$. Instead of using this propagator the largest MBLE which have been generated via ED are $N=22$ \cite{Luitz2015} and use the shift-and-invert technique;  this technique repeatedly applies $(H-\lambda)^{-1}$ to a random state converging to an eigenstate with eigenvalue close to $\lambda$.  This technique overcomes the problem of nearly degenerate eigenstates by inverting the spectrum. We develop an analogous approach in the MPS language called SIMPS.  Starting with a random MPS, we iteratively apply $(H - \lambda)^{-1}$ until we reach an MPS that well approximates an eigenstate close to $\lambda$.  The convergence speed of this method is geometric in the limit of large bond dimension, and mostly controlled by the ratio of the second dominant eigenvalue to the first dominant eigenvalue present in the state
\begin{equation}
\rho = \left| \frac{(E_2 - \lambda)^{-1}}{(E_1 - \lambda)^{-1}} \right| = \left| \frac{E_1 - \lambda}{E_2 - \lambda} \right| < 1
\end{equation}
where $E_1$ is the eigenvalue of $H$ closest to $\lambda$ and $E_2$ is the second closest to $\lambda$. As an example, given an MPS $|\psi \rangle = a|E_1\rangle + b |E_2\rangle + \cdots$ (where $\cdots$ means other states far away from $\lambda$), the MPS $|\varphi\rangle = (H-\lambda)^{-N}|\psi\rangle$ has an energy of
\begin{equation}
\langle E \rangle \approx \frac{E_1+E_2\frac{|b|^2}{|a|^2} \rho^{2N}}{1+\frac{|b|^2}{|a|^2} \rho^{2N}}
\label{convergence}
\end{equation}
Therefore, the energy of the MPS decays exponentially in $N$. To increase the convergence speed, one can fine tune $\lambda$ such that $\rho$ is much smaller than $1$, by fitting the energy trace of $\langle E \rangle$ vs. $N$ accumulated during a particular run to Eq.~\eqref{convergence}.  See Supplementary Figures for a prototypical example of SIMPS converging and an example of this decay with a poorly chosen value of $\lambda$. 

\begin{figure}[th]
\centering
\includegraphics[width=0.8\columnwidth]{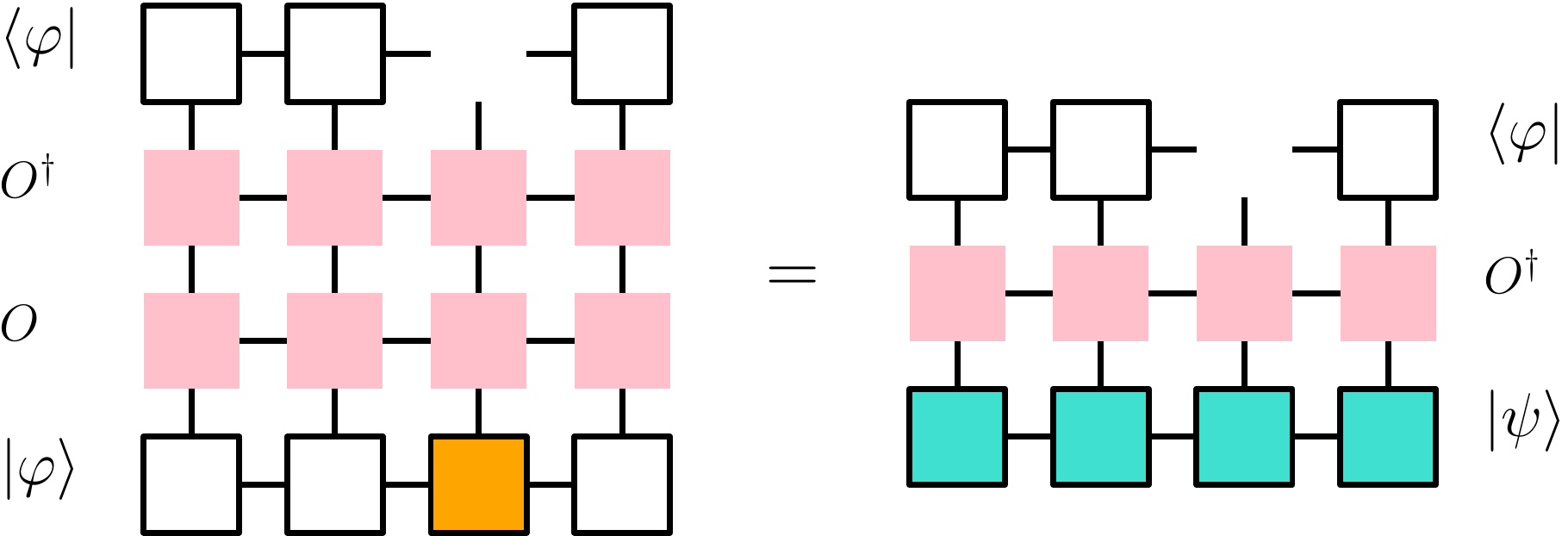}
\caption{Pictorial representation for optimizing $||O|\varphi\rangle-|\psi\rangle||^2$. Here as an example, we consider optimizing the third site (colored orange) for  a $L=4$ chain. The MPO $O$ (pink) and the MPS $|\psi\rangle$ (cyan) are given. To update the orange block, we solve a linear equation,  treating the orange block as an unknown vector $x$; the other part of the network on the left hand side, after being contracted, amounts to a symmetric semi positive-definite matrix $A$, and the matrices on the right hand side of the equation becomes a known vector $b$. The entire MPS is optimized by sweeping. Given an cutoff $\epsilon$, the sweeping can be stopped by checking if $|1-\langle \varphi |O^{\dagger}| \psi \rangle | <\epsilon$, which can be done without additional computational effort since $\langle \varphi |O^{\dagger}| \psi \rangle$ can be calculated by contracting the solved orange block with the rest of the network on the right.}
\label{invDMRG}
\end{figure}

\textbf{Inverse DMRG:} The key to SIMPS is the development of an inverse-DMRG approach which variationally applies the inverse of an MPO to an MPS.  Just like inverting a non-degenerate matrix $A$ directly and applying it to a vector $b$ is more difficult than solving $Ax=b$, inverting an MPO directly suffers two major problems -- (1) it lacks an efficient algorithm; (2) it may potentially require very large bond dimension for the inverted MPO. The more efficient alternative is to construct a MPS $|\varphi\rangle$ which variationally approaches $O^{-1} |\psi\rangle$, for a given MPO $O$ and MPS $|\psi\rangle$. To this end, we propose to build the MPS $|\varphi\rangle$ by minimizing $||O|\varphi\rangle-|\psi\rangle||^2$,
\begin{equation}
\frac{\partial}{\partial \varphi^*_{i,\sigma}} ||O|\varphi\rangle-|\psi\rangle||^2 = 0,
\end{equation}
where $i$ stands for a particular site and $\sigma$ is a label for degrees of freedom on site $i$. The above equation leads to the following variational prescription
\begin{equation}
\frac{\partial}{\partial \varphi^*_{i,\sigma}} \langle \varphi |O^{\dagger}O| \varphi \rangle = \frac{\partial}{\partial \varphi^*_{i,\sigma}} \langle \varphi |O^{\dagger}| \psi \rangle.
\end{equation}
which is described pictorially in Fig. \ref{invDMRG}. To update the matrices at a particular site, one needs to solve a linear equation problem that involves a dense, symmetric, and semi positive-definite matrix. To optimize the entire MPS $|\varphi\rangle$, one needs to sweep the chain back and forth as in standard DMRG.

Precision of this method is controlled by the bond dimension of the MPS $|\varphi\rangle$, and the number of sweeps. The computationally dominate piece is in solving the dense system of linear equations. Substantial speedup can be gained by using conserved quantum numbers (QN) (although we currently don't use this) which reduce the size of equations. The computational cost of inverse-DMRG scales as $O(L(pM^2)^3)$ for using a direct solver, and optimistically $O(L(pM^2)^2)$ when using an iterative solver, where $p$ is the number of physical degrees of freedom per site, and $M$ is the maximum bond dimension of the MPS.  We currently do direct solves up to $M=60$.

\begin{figure}[th]
\centering
\includegraphics[width=\columnwidth]{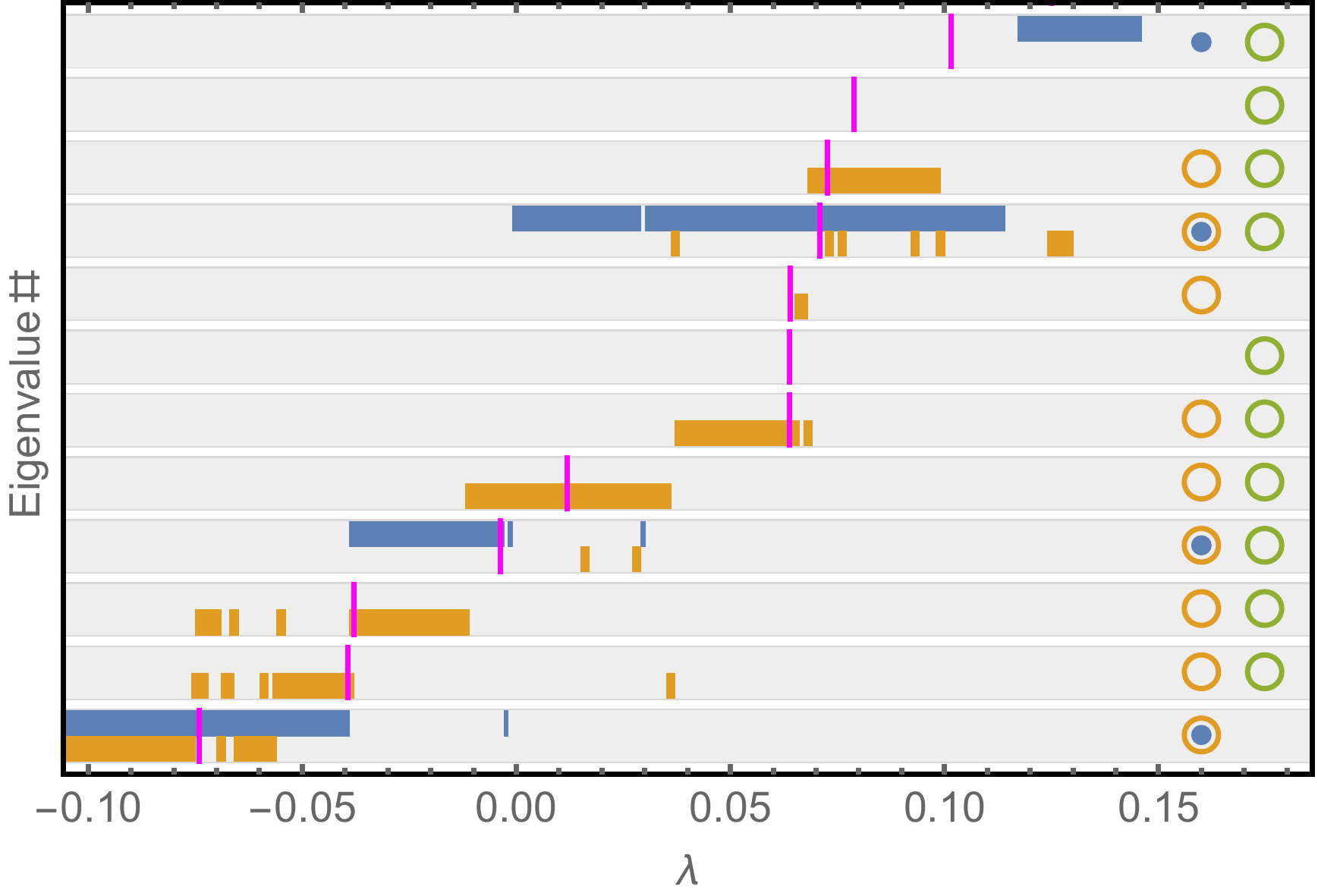}
\caption{Convergence to eigenstates 503-514. Magenta lines indicate the twelve eigenvalues in this range. SIMPS: Blue and yellow bars indicate the basins of convergence, as a function of parameter $\lambda$, found for two random MPS initial conditions. Yellow and blue dots indicate whether sweeps of the two respective initial conditions find a given eigenvalue. ES-DMRG: Green dots indicate whether ES-DMRG found the eigenvalue after being run with 6144 random product state initial conditions.
}
\label{L10_W8_SIMPS}
\end{figure}

\begin{figure}[th]
\centering
\includegraphics[width=\columnwidth]{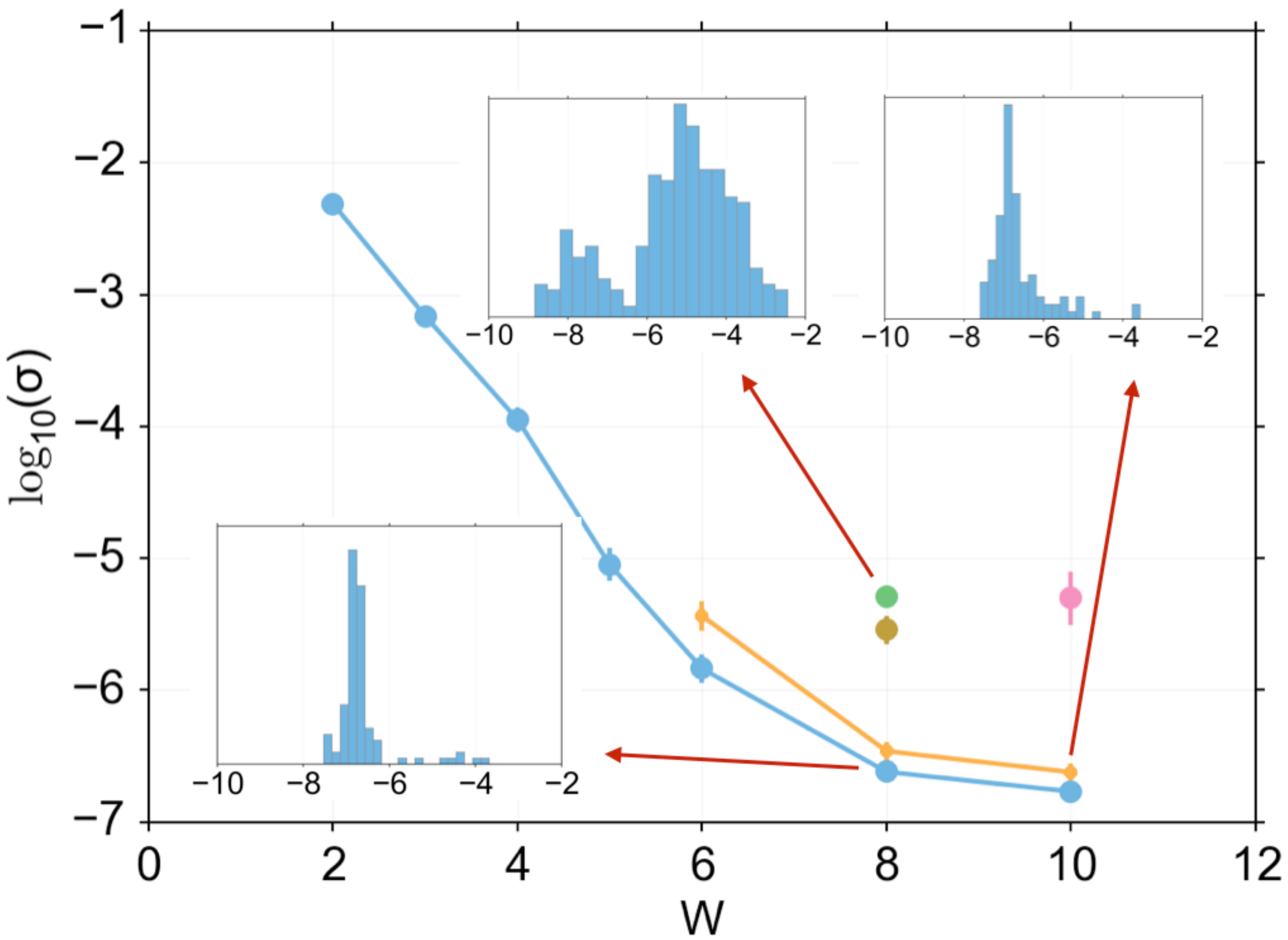}
\caption{Values of $\langle \log_{10}(\sigma) \rangle$ for various algorithms and parameters at different disorder strengths $W$.  Blue line ($L=30$) and orange line ($L=40$) are SIMPS runs at $M=60$ with $\lambda$ tuned exactly to the middle of the spectrum for each disorder realization, where the lowest and highest eigenvalues are obtained via DMRG and the disorder configurations are the same scaled pattern at different $W$ for $L=30$ and $L=40$, respectively.  Values of $W \le 5$ are not necessarily reliably converged.  Green point is ES-DMRG at $M=20$, $L=30$ and pink point is ES-DMRG at $M=20$, $L=100$. The dark green point is SIMPS at $M=20$, $L=30$ showing that ES-DMRG and SIMPS converge to similar accuracy at the same bond dimension and chain length. Insets are distributions of $\sigma$ at the respective parameters.}
\label{std_Hist}
\end{figure}

\textbf{Generating Eigenstates:} We start by producing eigenstates at length $L=10$ such that they can be compared with ED. Here, for both algorithms, we consider a fixed disorder configuration  with $W=8$ and artificially limit bond dimension to $M=12$. Since ES-DMRG can't target particular energies, we run the algorithm many times verifying that the energies it finds match those of the true Hamiltonian (see Fig.~\ref{L10_W8_SIMPS}).  In SIMPS, we have a tunable shift parameter $\lambda$ and focus on the energy window of $\lambda \in [-0.1, 0.1]$, within which there are 12 eigenstates as shown by ED. 
When running at a limited bond dimension with fixed initial conditions, the SIMPS algorithm does not always hit the targeted eigenstate. Remarkably, it does find an eigenstate near the desired energy with high fidelity.  In fact, if we consider 
\begin{equation}
P(\lambda)  = \min_i \left[ 1- |\langle MPS(\lambda) | ED(i)  \rangle |\right]
\label{Fidelity_1}
\end{equation}
then for $L=10, M=12$ we never see a situation where the $P(\lambda) >10^{-5}$. Fig.~\ref{L10_W8_SIMPS} shows the eigenvalues identified during a $\lambda$ sweep for different initial conditions and bond-dimensions of SIMPS.  Notice that changing the initial conditions  allows targeting of eigenstates that were inaccessible with the original initial condition. Hence, using just two initial conditions it is possible to target 10 out of the 12 eigenstates. 

Another useful metric to quantify the quality of eigenstates, that can be applied to system sizes greater then those accessible to ED, is the standard deviation of the energy $\sigma=\sqrt{\langle H\rangle^2-\langle H^2\rangle}$. Note that while a true eigenstate has $\sigma=0$, numerically calculated standard deviations are limited by machine precision. Unsurprisingly, the high fidelity we see on short chains ($L=10, M=12$) corresponds to $\langle \log_{10} \sigma \rangle \approx -6.1$ which is beyond machine precision. 

\begin{figure}
\includegraphics[width=0.8\columnwidth]{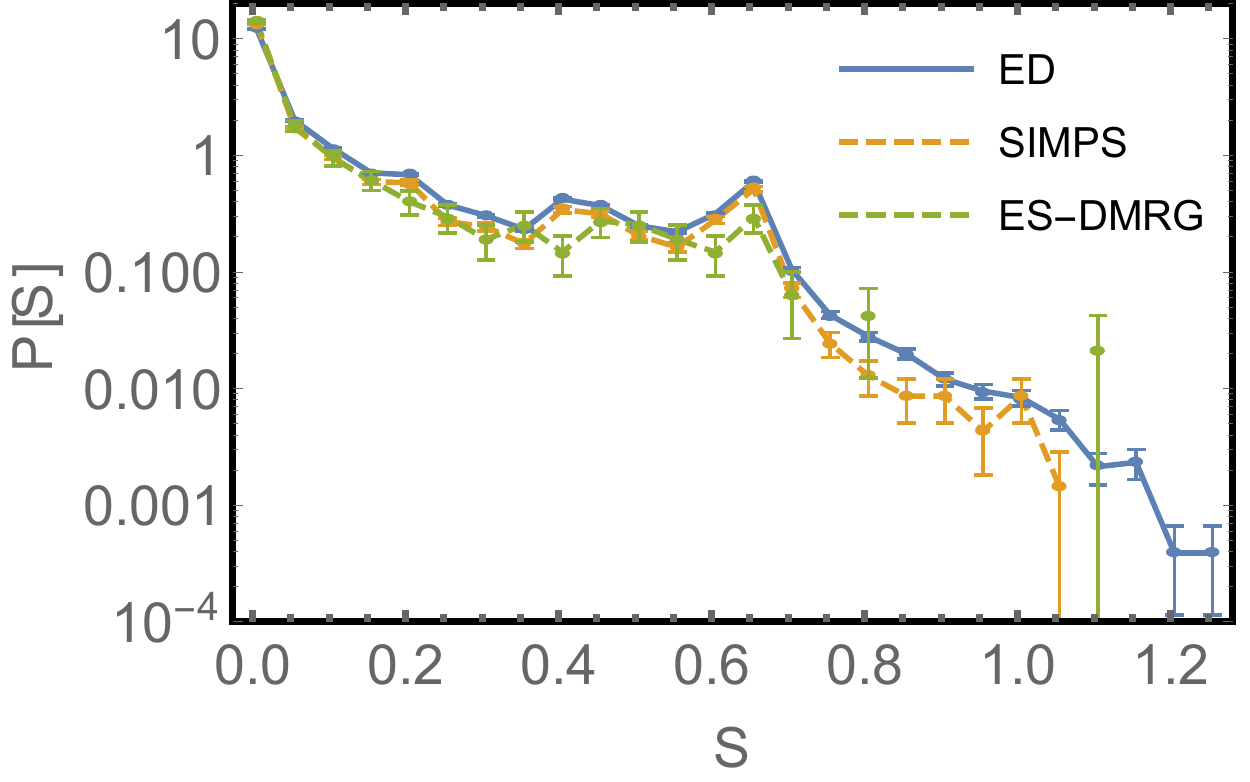}
\caption{
A comparison of mid-bond entanglement entropy histograms generated using ED, SIMPS, and ES-DMRG eigenstates. To generate the histograms, we fixed the disorder strength at $W=8$, chain length at $L=10$, and the maximum bond dimension used in SIMPS and ES-DMRG at $M=12$. Next, we generated 101 disorder realizations for ED and SIMPS (ES-DMRG data is for 320 different disorder realizations). Using these we construct all 103,424 eigenstates using ED, 14,017 eigenstates using SIMPS, and 951 eigenstates using ES-DMRG. Error bars  correspond to points $n_i \pm \sqrt{n_i}$, where $n_i$ is the number of entanglement entropies in the $i$-th bin. The error bars should be treated as a lower bound as they do not take into account the fact that for the same disorder configuration the mid-bond entanglement entropy of different eigenstates is correlated, and hence the number of independent samples is lower than what is expected by the naive estimate above. 
}
\label{fig:hist}
\end{figure}

As both SIMPS and ES-DMRG algorithms stochastically converge to eigenstates, one has to check whether the states that our algorithms converge to are typical or  biased. Specifically, we want to verify there isn't a bias toward eigenstates with lower entanglement. To verify this is not occurring we compare histograms of entanglement entropies for eigenstates generated using ED to those generated using SIMPS and ES-DMRG at artificially reduced bond-dimension M=12 (see Fig.~\ref{fig:hist}). From the comparison we observe that both SIMPS and ES-DMRG are sampling states at a given entanglement with the same frequency as ED and hence there is no systematic bias.

\begin{figure}[th]
\centering
\includegraphics[width=0.8\columnwidth]{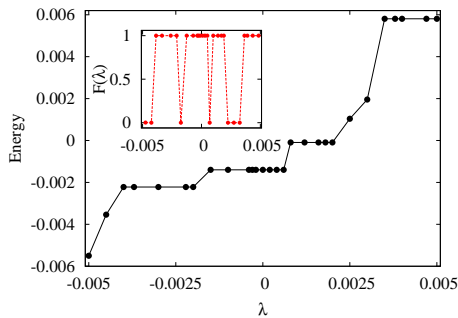}
\caption{Energy vs $\lambda$ for $L=30, W=10$. 29 MPSs obtained using SIMPS with $M=40$ are displayed on this plot, with 8 of them having distinct energies. Inset: Fidelity function $F(\lambda)$ as define in Eq.~\eqref{Fidelity_2}. $F(\lambda)$ also shows 8 distinct states, and matches exactly with the energy curve.}
\label{L30_M40}
\end{figure}

We find that, using SIMPS at $M=60$, we can obtain eigenstates of chains of length $L=30,40$ and $W=8$ to average values of  $\langle \sigma \rangle$ below machine precision and with a distribution that is largely peaked at machine precision (see fig.~\ref{L10_W8_SIMPS}).  These tests are conducted for the worst case scenario -- we target $\lambda$ exactly to the middle of the energy spectra (lowest and highest energies calculated by DMRG) where the many-body density of states is near maximum. Running ES-DMRG at $M=20$ for chains with \{$L=30$, $W=8$\} and \{$L=100$, $W=10$\}, we find approximate eigenstates with $\langle \log_{10}(\sigma) \rangle$ somewhat larger than SIMPS. 

Of course, for long chains, the inter-level spacing is going to be smaller then the $\sigma$ which can be resolved by machine precision and so it is useful to have an additional measure to check that we are not converging to a linear combination of eigenstates.  It is expected that when tuning $\lambda$ with a small increment, two successive MPSs built by the method should either be the same eigenstate or be orthogonal to each other. This means the fidelity
\begin{equation}
F(\lambda_i) = \left| \langle MPS(\lambda_i) | MPS(\lambda_{i+1}) \rangle \right|
\label{Fidelity_2}
\end{equation}
should alternate between zero and one. If instead, each independent simulation at $\lambda$ produced a different linear superposition of nearby eigenstates we would expect $F(\lambda_i)$ to vary continuously between zero and one. The inset of  Fig.~\ref{L30_M40} shows eight distinct eigenstates with $F(\lambda_i)$ alternating between 1 (the same state) or 0 (an orthogonal state).

\textbf{Many-body localization in large systems:} We now use the tools that we have developed to test, in a previously inaccessible regime of long chains, three key properties of MBL matter: failure to thermalize, low entanglement entropy of highly excited eigenstates, and a large number of local excitations.

\begin{figure}[th]
\centering
\includegraphics[width=\columnwidth]{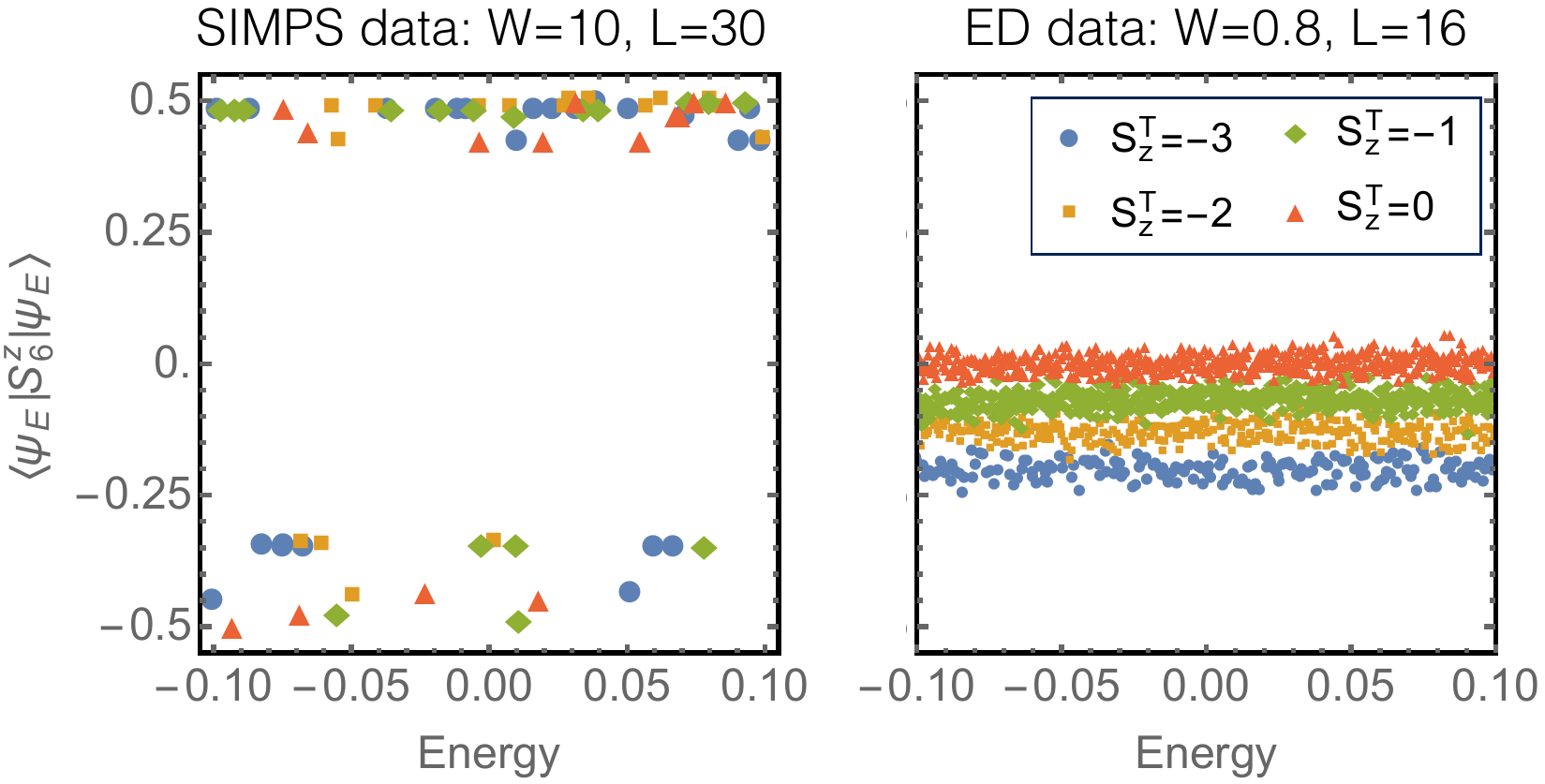}
\caption{\textit{Left:} Site value of $\langle S^6_z \rangle$  as a function of energy at $L=30$ for a fixed disorder configuration with disorder strength $W=10$. Eigenstates are obtained using SIMPS, with $M=40$. Note the small energy range. The energy scan was performed with 200 $\lambda$ evenly spaced in $[-0.1,0.1]$. In the $S^T_z=-3,-2,-1,0$ sectors, a total of 74 distinct MPSs were found. \textit{Right:} Site value of $\langle S^6_z \rangle $ for $L=16, W=0.8$ generated using ED.} 
\label{L30_M40_ETH}
\end{figure}

The eigenstate thermalization hypothesis (ETH) states that in thermalized quantum systems nearby eigenstates have very similar expectation values of local observables~\cite{Deutsch1991,Srednicki1994,Rigol2008}. We test this by computing the expectation value of a representative local observable $\langle S^z_6 \rangle$ for two chains: one with $L=16$ at $W=0.8$ (expected to obey ETH) and another with $L=30$ at $W=8$ (expected to violate ETH). Specifically, we choose eigenstates from a \textit{small energy window} in the range $[-0.1,0.1]$. We also filter the eigenstates by total $S_z^T=\sum_i S^z_i$. For the case $W=8$, SIMPS discovered 74 eigenstates that were close in energy (but not consecutive). As shown in Fig.~\ref{L30_M40_ETH}, the corresponding $\langle S^z_6 \rangle$ vary wildly confirming the violation of ETH. On the other hand, for $W=0.8$ ETH prevails -- $\langle S^z_6 \rangle$ lies in a narrow band, with the position of the band depending on the global conserved quantity $S_z^T$ [see Fig.~\ref{L30_M40_ETH}]. 

\begin{figure}[th]
\centering
\includegraphics[width=\columnwidth]{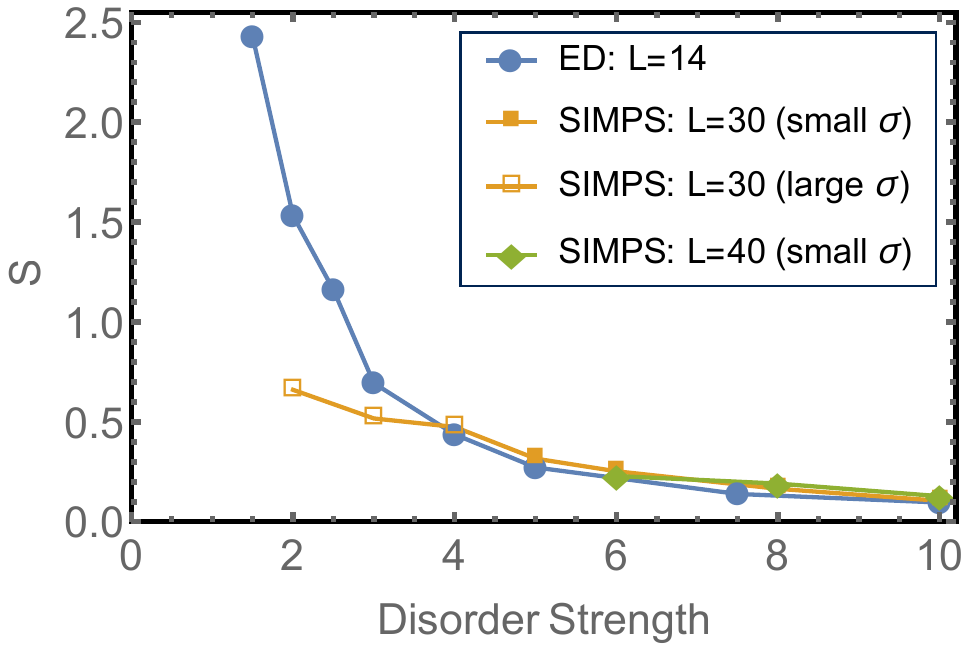}
\caption{Dependence of mid-bond entanglement entropy $S$ on the disorder strength for various chain lengths. 50 disorder realizations were used for  $L=14$ and 100 for $L=30$ and $L=40$. Open markers indicate points at low disorder strengths for which SIMPS failed to obtain states with low standard deviation of the energy. For SIMPS we set $M=60$ and tuned $\lambda$ to the exact middle of the eigenspectrum for each disorder realization.}
\label{S_Hist}
\end{figure}

The second key property that we investigate is the mid-bond entanglement entropy of highly excited eigenstates. In the ergodic phase these eigenstates have extensive mid-bond entanglement entropy while in the MBL phase the mid-bond entanglement entropy should saturate as a function of system size. In Fig.~\ref{L30_M40_ETH} we plot the mean mid-bond entanglement entropy as a function of disorder strength for chains of length $L=14$ (ED) and $L=30,40$ (SIMPS). We observe that for $W>4$ the mid-bond entanglement entropy is essentially independent of systems size, thus confirming the predicted saturation in the MBL phase. On the other hand for $W<4$ the ED and SIMPS data strongly disagree. We attribute this disagreement to the failure of SIMPS to find high quality eigenstates in this near-ergodic and ergodic regime as demonstrated in Fig.~\ref{std_Hist}. 

\begin{figure}[th]
\centering
\includegraphics[width=\columnwidth]{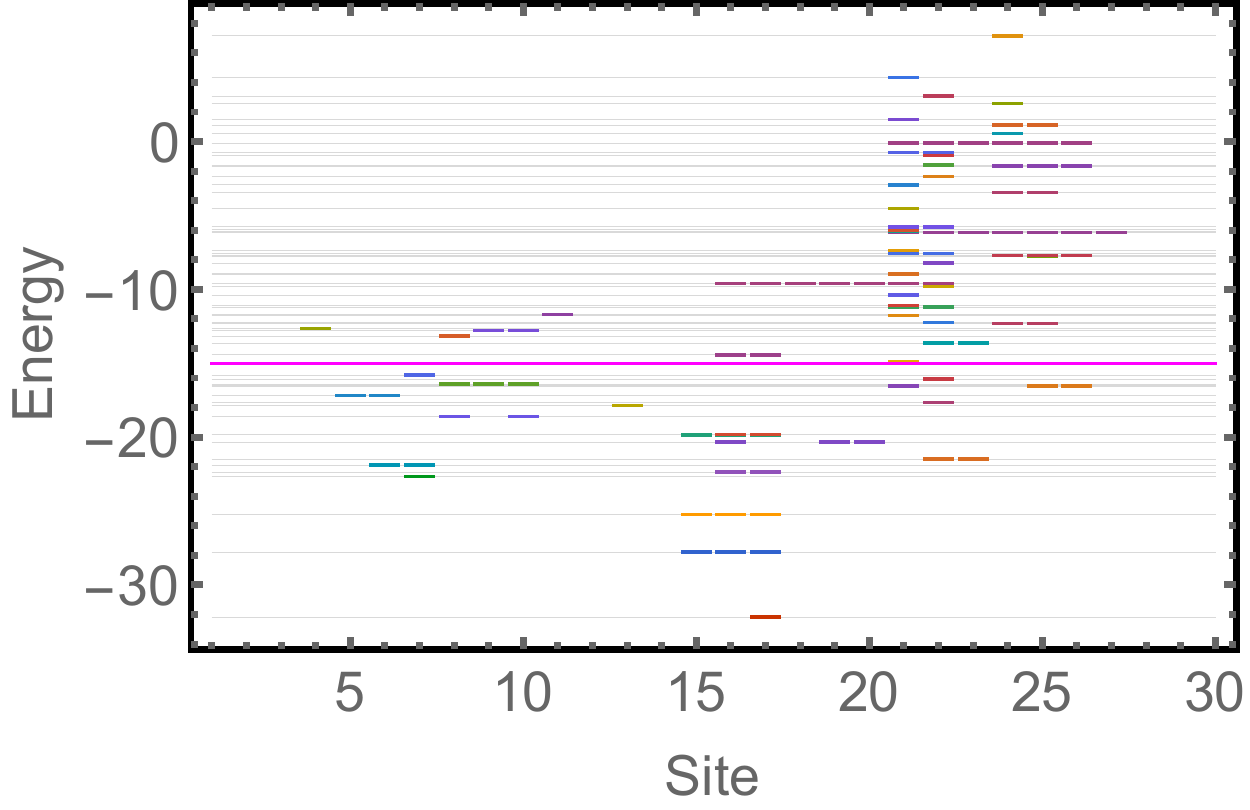}
\caption{Spectrum of 62 excitations (colored bars) produced by local modifications of a reference MPS eigenstate (long magenta line). The horizontal axis marks the site on which the MPS is being modified. Bars of the same color correspond to the same excitation as determined by their energies. [$L=30$, $W=10$, $M=15$, excited eigenstates all have $\sigma<0.0005$].} 
\label{L30_M15_LX}
\end{figure}

The third property that we investigate is the existence of a large number of local excitations of the system. In Fig.~\ref{L30_M15_LX} we show a spectrum of these local excitations constructed with respect to an eigenstate at $E=-15.002$. To construct a local excitation on a given site, we take the following steps. (1) We bring the reference state to canonical form with respect to the target site. (2) We obtain the effective Hamiltonian on the target site. (3) We construct an MPS for {\it all} eigenstates of the effective Hamiltonian on the target site. (4) Most of the MPSs produced in step (3) are superpositions of a large number of local excitations, and hence are not necessarily good eigenstates. Therefore, we retain only the MPSs that have a low $\sigma$. Surprisingly, even after the filtering step, we find a large number of local excitations (there are 62 distinct ones in Fig.~\ref{L30_M15_LX}). 

\textbf{Outlook:} In our manuscript, we have presented two algorithms ES-DMRG and SIMPS for finding matrix product state representations of inner eigenstates. We demonstrated the functionality and efficiency of both algorithms by showing that they can be used to produce good variational MPS approximations of the excited states of a fully many-body localized Hamiltonian in one dimension. We have also used these algorithms to demonstrate three key attributes of many-body localization, the breakdown of the eigenstate thermalization hypothesis, the saturation of the entanglement entropy, and the presence of a large number of local excitations, in a previously inaccessible regime of long chains, $L=30,40$. 

We are particularly excited about two potential consequences of our work for many-body localization. First, as demonstrated in Fig.~\ref{S_Hist}, the algorithms that we have developed seem to work right up to the localization-delocalization transition. We speculate that with further improvements these algorithms can be used to probe the transition from the localized side~\cite{Grover2014}. Second, it seems possible that by collating and orthogonalizing the ``local excitation'' data of Fig.~\ref{L30_M15_LX} one can construct and characterize the local integrals of motion that are central the many-body localization phenomenology. 

We believe that the methods that we have developed are not limited to the study of strongly-localized matter. Specifically, these methods have the potential to be significantly better for finding low-lying excited states of conventional Hamiltonians as compared to current state of the art methods -- an extremely important problem both in quantum chemistry and condensed matter physics. Moreover, we suspect that further improvements can be made to our algorithms by applying more sophisticated diagonalization methods to matrix product states.

\textbf{Note Added:} During the preparation of this manuscript we became aware of two independent works that also developed an algorithm similar to ES-DMRG for finding excited eigenstates in MBL systems~\cite{Khemani2015} and molecular systems~\cite{hu2015excited}. 

\textbf{Acknowledgements: } The authors thank Hitesh J. Changlani, Victor Chua, David Huse, Vedika Khemani, Vadim Oganesyan, Shivaji Sondhi, and Gil Refael for helpful discussions. We acknowledge MPI-PKS and the Aspen Center for Physics for their hospitality.  This research is part of the Blue Waters sustained-petascale computing project, which is supported by the National Science Foundation (award number ACI 1238993) and the state of Illinois. Blue Waters is a joint effort of the University of Illinois at Urbana-Champaign and its National Center for Supercomputing Applications. DP acknowledges support from the Charles E. Kaufman foundation; BKC and XY acknowledges support from grant DOE, SciDAC FG02-12ER46875.

\section{Supplementary figures}
\begin{figure}[th]
\centering
\includegraphics[scale=.30]{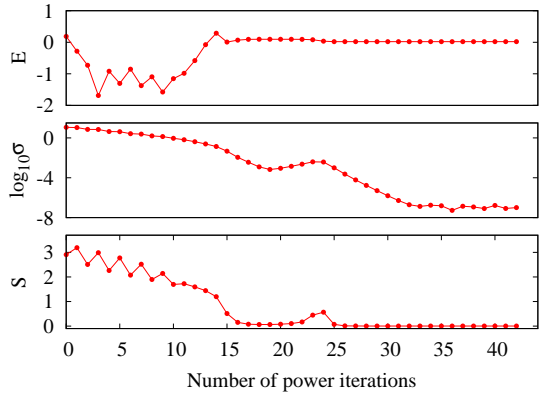}
\caption{Convergence behaviors of one SIMPS run for $L=30$, $W=8$ and $M=40$. Initial MPS, at iteration 0, is a random, normalized MPS. Top to bottom: energy $E$, log of standard deviation $\log_{10}(\sigma)$, mid-bond von Neumann entanglement entropy $S$. In inverse-DMRG, when starting the sweeping for $|\varphi\rangle = (H-\lambda)^{-1} |\psi\rangle$, one could (a) start with $|\varphi\rangle$ the same as $|\psi\rangle$; (b) start with random $|\varphi\rangle$. The second approach can help to overcome local minima during inverse-DMRG sweeping. In the plot above, $|\varphi\rangle$ was set to a random MPS at iteration 1 and 20.}
\label{M32_invDMRG}
\end{figure}		

\begin{figure}[th]
\centering
\includegraphics[scale=.30]{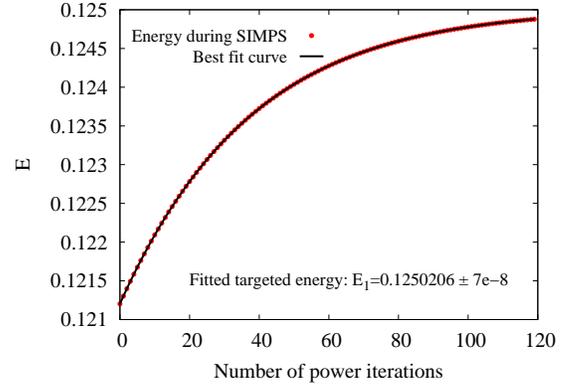}
\caption{A slow converging case of SIMPS for $L=10$, $W=8$ and $M=12$ with a poorly chosen initial $\lambda$. After 240 power iterations, the MPS has not yet converged (only the last 120 iterations are shown and fitted). To fine tune $\lambda$, one can fit the energy versus number of iterations to a function taking the form of Eq. \eqref{convergence}. The results in many cases are useful in learning how to tune $\lambda$ (setting $\lambda$ as the fitted targeted energy $E_1$ is an option). For the case displayed above, $E_1$ to high precision is already close to the ED eigenvalue 0.125020071154031.}
\label{Fine_Tuning}
\end{figure}

\bibliography{MPS}

\begin{thebibliography}{39}%
\makeatletter
\providecommand \@ifxundefined [1]{%
 \@ifx{#1\undefined}
}%
\providecommand \@ifnum [1]{%
 \ifnum #1\expandafter \@firstoftwo
 \else \expandafter \@secondoftwo
 \fi
}%
\providecommand \@ifx [1]{%
 \ifx #1\expandafter \@firstoftwo
 \else \expandafter \@secondoftwo
 \fi
}%
\providecommand \natexlab [1]{#1}%
\providecommand \enquote  [1]{``#1''}%
\providecommand \bibnamefont  [1]{#1}%
\providecommand \bibfnamefont [1]{#1}%
\providecommand \citenamefont [1]{#1}%
\providecommand \href@noop [0]{\@secondoftwo}%
\providecommand \href [0]{\begingroup \@sanitize@url \@href}%
\providecommand \@href[1]{\@@startlink{#1}\@@href}%
\providecommand \@@href[1]{\endgroup#1\@@endlink}%
\providecommand \@sanitize@url [0]{\catcode `\\12\catcode `\$12\catcode
  `\&12\catcode `\#12\catcode `\^12\catcode `\_12\catcode `\%12\relax}%
\providecommand \@@startlink[1]{}%
\providecommand \@@endlink[0]{}%
\providecommand \url  [0]{\begingroup\@sanitize@url \@url }%
\providecommand \@url [1]{\endgroup\@href {#1}{\urlprefix }}%
\providecommand \urlprefix  [0]{URL }%
\providecommand \Eprint [0]{\href }%
\providecommand \doibase [0]{http://dx.doi.org/}%
\providecommand \selectlanguage [0]{\@gobble}%
\providecommand \bibinfo  [0]{\@secondoftwo}%
\providecommand \bibfield  [0]{\@secondoftwo}%
\providecommand \translation [1]{[#1]}%
\providecommand \BibitemOpen [0]{}%
\providecommand \bibitemStop [0]{}%
\providecommand \bibitemNoStop [0]{.\EOS\space}%
\providecommand \EOS [0]{\spacefactor3000\relax}%
\providecommand \BibitemShut  [1]{\csname bibitem#1\endcsname}%
\let\auto@bib@innerbib\@empty
\bibitem [{\citenamefont {Basko}\ \emph {et~al.}(2006)\citenamefont {Basko},
  \citenamefont {Aleiner},\ and\ \citenamefont {Altshuler}}]{Basko2006}%
  \BibitemOpen
  \bibfield  {author} {\bibinfo {author} {\bibfnamefont {D.~M.}\ \bibnamefont
  {Basko}}, \bibinfo {author} {\bibfnamefont {I.~L.}\ \bibnamefont {Aleiner}},
  \ and\ \bibinfo {author} {\bibfnamefont {B.~L.}\ \bibnamefont {Altshuler}},\
  }\href {http://linkinghub.elsevier.com/retrieve/pii/S0003491605002630}
  {\bibfield  {journal} {\bibinfo  {journal} {Annals of Physics}\ }\textbf
  {\bibinfo {volume} {321}},\ \bibinfo {pages} {1126} (\bibinfo {year}
  {2006})}\BibitemShut {NoStop}%
\bibitem [{\citenamefont {Basko}\ \emph {et~al.}(2007)\citenamefont {Basko},
  \citenamefont {Aleiner},\ and\ \citenamefont {Altshuler}}]{Basko2007}%
  \BibitemOpen
  \bibfield  {author} {\bibinfo {author} {\bibfnamefont {D.~M.}\ \bibnamefont
  {Basko}}, \bibinfo {author} {\bibfnamefont {I.~L.}\ \bibnamefont {Aleiner}},
  \ and\ \bibinfo {author} {\bibfnamefont {B.~L.}\ \bibnamefont {Altshuler}},\
  }\href {\doibase 10.1103/PhysRevB.76.052203} {\bibfield  {journal} {\bibinfo
  {journal} {Phys. Rev. B}\ }\textbf {\bibinfo {volume} {76}},\ \bibinfo
  {pages} {052203} (\bibinfo {year} {2007})}\BibitemShut {NoStop}%
\bibitem [{\citenamefont {Oganesyan}\ and\ \citenamefont
  {Huse}(2007)}]{Oganesyan2007}%
  \BibitemOpen
  \bibfield  {author} {\bibinfo {author} {\bibfnamefont {V.}~\bibnamefont
  {Oganesyan}}\ and\ \bibinfo {author} {\bibfnamefont {D.~A.}\ \bibnamefont
  {Huse}},\ }\href {\doibase 10.1103/PhysRevB.75.155111} {\bibfield  {journal}
  {\bibinfo  {journal} {Phys. Rev. B}\ }\textbf {\bibinfo {volume} {75}},\
  \bibinfo {pages} {155111} (\bibinfo {year} {2007})}\BibitemShut {NoStop}%
\bibitem [{\citenamefont {Pal}\ and\ \citenamefont {Huse}(2010)}]{Pal2010}%
  \BibitemOpen
  \bibfield  {author} {\bibinfo {author} {\bibfnamefont {A.}~\bibnamefont
  {Pal}}\ and\ \bibinfo {author} {\bibfnamefont {D.~A.}\ \bibnamefont {Huse}},\
  }\href {\doibase 10.1103/PhysRevB.82.174411} {\bibfield  {journal} {\bibinfo
  {journal} {Phys. Rev. B}\ }\textbf {\bibinfo {volume} {82}},\ \bibinfo
  {pages} {174411} (\bibinfo {year} {2010})}\BibitemShut {NoStop}%
\bibitem [{\citenamefont {Aleiner}\ \emph {et~al.}(2010)\citenamefont
  {Aleiner}, \citenamefont {Altshuler},\ and\ \citenamefont
  {Shlyapnikov}}]{Aleiner2010}%
  \BibitemOpen
  \bibfield  {author} {\bibinfo {author} {\bibfnamefont {I.~L.}\ \bibnamefont
  {Aleiner}}, \bibinfo {author} {\bibfnamefont {B.~L.}\ \bibnamefont
  {Altshuler}}, \ and\ \bibinfo {author} {\bibfnamefont {G.~V.}\ \bibnamefont
  {Shlyapnikov}},\ }\href@noop {} {\bibfield  {journal} {\bibinfo  {journal}
  {Nat. Phys.}\ }\textbf {\bibinfo {volume} {6}},\ \bibinfo {pages} {900}
  (\bibinfo {year} {2010})}\BibitemShut {NoStop}%
\bibitem [{\citenamefont {Monthus}\ and\ \citenamefont
  {Garel}(2010)}]{Monthus2010}%
  \BibitemOpen
  \bibfield  {author} {\bibinfo {author} {\bibfnamefont {C.}~\bibnamefont
  {Monthus}}\ and\ \bibinfo {author} {\bibfnamefont {T.}~\bibnamefont
  {Garel}},\ }\href {\doibase 10.1103/PhysRevB.81.134202} {\bibfield  {journal}
  {\bibinfo  {journal} {Phys. Rev. B}\ }\textbf {\bibinfo {volume} {81}},\
  \bibinfo {pages} {134202} (\bibinfo {year} {2010})}\BibitemShut {NoStop}%
\bibitem [{\citenamefont {Imbrie}(2014)}]{Imbrie2014}%
  \BibitemOpen
  \bibfield  {author} {\bibinfo {author} {\bibfnamefont {J.~Z.}\ \bibnamefont
  {Imbrie}},\ }\href {http://www.arxiv.org/abs/1403.7837} {\enquote {\bibinfo
  {title} {On many-body localization for quantum spin chains},}\ } (\bibinfo
  {year} {2014}),\ \Eprint {http://arxiv.org/abs/1403.7837} {1403.7837}
  \BibitemShut {NoStop}%
\bibitem [{\citenamefont {Nandkishore}\ and\ \citenamefont
  {Huse}(2015)}]{Nandkishore2015}%
  \BibitemOpen
  \bibfield  {author} {\bibinfo {author} {\bibfnamefont {R.}~\bibnamefont
  {Nandkishore}}\ and\ \bibinfo {author} {\bibfnamefont {D.~A.}\ \bibnamefont
  {Huse}},\ }\href@noop {} {\bibfield  {journal} {\bibinfo  {journal} {Annual
  Review of Condensed Matter Physics}\ }\textbf {\bibinfo {volume} {6}},\
  \bibinfo {pages} {15} (\bibinfo {year} {2015})}\BibitemShut {NoStop}%
\bibitem [{\citenamefont {Bauer}\ and\ \citenamefont
  {Nayak}(2013)}]{Bauer2013}%
  \BibitemOpen
  \bibfield  {author} {\bibinfo {author} {\bibfnamefont {B.}~\bibnamefont
  {Bauer}}\ and\ \bibinfo {author} {\bibfnamefont {C.}~\bibnamefont {Nayak}},\
  }\href {http://stacks.iop.org/1742-5468/2013/i=09/a=P09005} {\bibfield
  {journal} {\bibinfo  {journal} {Journal of Statistical Mechanics: Theory and
  Experiment}\ }\textbf {\bibinfo {volume} {2013}},\ \bibinfo {pages} {P09005}
  (\bibinfo {year} {2013})}\BibitemShut {NoStop}%
\bibitem [{\citenamefont {Serbyn}\ \emph
  {et~al.}(2013{\natexlab{a}})\citenamefont {Serbyn}, \citenamefont
  {Papi\'{c}},\ and\ \citenamefont {Abanin}}]{Serbyn2013}%
  \BibitemOpen
  \bibfield  {author} {\bibinfo {author} {\bibfnamefont {M.}~\bibnamefont
  {Serbyn}}, \bibinfo {author} {\bibfnamefont {Z.}~\bibnamefont {Papi\'{c}}}, \
  and\ \bibinfo {author} {\bibfnamefont {D.~A.}\ \bibnamefont {Abanin}},\
  }\href@noop {} {\bibfield  {journal} {\bibinfo  {journal} {Phys. Rev. Lett.}\
  }\textbf {\bibinfo {volume} {111}},\ \bibinfo {pages} {127201} (\bibinfo
  {year} {2013}{\natexlab{a}})}\BibitemShut {NoStop}%
\bibitem [{\citenamefont {Huse}\ \emph {et~al.}(2014)\citenamefont {Huse},
  \citenamefont {Nandkishore},\ and\ \citenamefont {Oganesyan}}]{Huse2014}%
  \BibitemOpen
  \bibfield  {author} {\bibinfo {author} {\bibfnamefont {D.~A.}\ \bibnamefont
  {Huse}}, \bibinfo {author} {\bibfnamefont {R.}~\bibnamefont {Nandkishore}}, \
  and\ \bibinfo {author} {\bibfnamefont {V.}~\bibnamefont {Oganesyan}},\ }\href
  {\doibase 10.1103/PhysRevB.90.174202} {\bibfield  {journal} {\bibinfo
  {journal} {Phys. Rev. B}\ }\textbf {\bibinfo {volume} {90}},\ \bibinfo
  {pages} {174202} (\bibinfo {year} {2014})}\BibitemShut {NoStop}%
\bibitem [{\citenamefont {Ros}\ \emph {et~al.}(2015)\citenamefont {Ros},
  \citenamefont {M{\"u}ller},\ and\ \citenamefont {Scardicchio}}]{Ros2015}%
  \BibitemOpen
  \bibfield  {author} {\bibinfo {author} {\bibfnamefont {V.}~\bibnamefont
  {Ros}}, \bibinfo {author} {\bibfnamefont {M.}~\bibnamefont {M{\"u}ller}}, \
  and\ \bibinfo {author} {\bibfnamefont {A.}~\bibnamefont {Scardicchio}},\
  }\href {\doibase http://dx.doi.org/10.1016/j.nuclphysb.2014.12.014}
  {\bibfield  {journal} {\bibinfo  {journal} {Nuclear Physics B}\ }\textbf
  {\bibinfo {volume} {891}},\ \bibinfo {pages} {420 } (\bibinfo {year}
  {2015})}\BibitemShut {NoStop}%
\bibitem [{\citenamefont {Vosk}\ and\ \citenamefont {Altman}(2013)}]{Vosk2013}%
  \BibitemOpen
  \bibfield  {author} {\bibinfo {author} {\bibfnamefont {R.}~\bibnamefont
  {Vosk}}\ and\ \bibinfo {author} {\bibfnamefont {E.}~\bibnamefont {Altman}},\
  }\href {\doibase 10.1103/PhysRevLett.110.067204} {\bibfield  {journal}
  {\bibinfo  {journal} {Phys. Rev. Lett.}\ }\textbf {\bibinfo {volume} {110}},\
  \bibinfo {pages} {067204} (\bibinfo {year} {2013})}\BibitemShut {NoStop}%
\bibitem [{\citenamefont {Vosk}\ and\ \citenamefont {Altman}(2014)}]{Vosk2014}%
  \BibitemOpen
  \bibfield  {author} {\bibinfo {author} {\bibfnamefont {R.}~\bibnamefont
  {Vosk}}\ and\ \bibinfo {author} {\bibfnamefont {E.}~\bibnamefont {Altman}},\
  }\href {\doibase 10.1103/PhysRevLett.112.217204} {\bibfield  {journal}
  {\bibinfo  {journal} {Phys. Rev. Lett.}\ }\textbf {\bibinfo {volume} {112}},\
  \bibinfo {pages} {217204} (\bibinfo {year} {2014})}\BibitemShut {NoStop}%
\bibitem [{\citenamefont {Pekker}\ \emph {et~al.}(2014)\citenamefont {Pekker},
  \citenamefont {Refael}, \citenamefont {Altman}, \citenamefont {Demler},\ and\
  \citenamefont {Oganesyan}}]{Pekker2014}%
  \BibitemOpen
  \bibfield  {author} {\bibinfo {author} {\bibfnamefont {D.}~\bibnamefont
  {Pekker}}, \bibinfo {author} {\bibfnamefont {G.}~\bibnamefont {Refael}},
  \bibinfo {author} {\bibfnamefont {E.}~\bibnamefont {Altman}}, \bibinfo
  {author} {\bibfnamefont {E.}~\bibnamefont {Demler}}, \ and\ \bibinfo {author}
  {\bibfnamefont {V.}~\bibnamefont {Oganesyan}},\ }\href {\doibase
  10.1103/PhysRevX.4.011052} {\bibfield  {journal} {\bibinfo  {journal} {Phys.
  Rev. X}\ }\textbf {\bibinfo {volume} {4}},\ \bibinfo {pages} {011052}
  (\bibinfo {year} {2014})}\BibitemShut {NoStop}%
\bibitem [{\citenamefont {Vosk}\ \emph {et~al.}(2015)\citenamefont {Vosk},
  \citenamefont {Huse},\ and\ \citenamefont {Altman}}]{Vosk2015}%
  \BibitemOpen
  \bibfield  {author} {\bibinfo {author} {\bibfnamefont {R.}~\bibnamefont
  {Vosk}}, \bibinfo {author} {\bibfnamefont {D.~A.}\ \bibnamefont {Huse}}, \
  and\ \bibinfo {author} {\bibfnamefont {E.}~\bibnamefont {Altman}},\ }\href
  {http://www.arxiv.org/abs/1412.3117} {\enquote {\bibinfo {title} {Theory of
  the many-body localization transition in one dimensional systems},}\ }
  (\bibinfo {year} {2015}),\ \Eprint {http://arxiv.org/abs/1412.3117}
  {1412.3117} \BibitemShut {NoStop}%
\bibitem [{\citenamefont {Potter}\ \emph {et~al.}(2015)\citenamefont {Potter},
  \citenamefont {Vasseur},\ and\ \citenamefont {Parameswaran}}]{Potter2015}%
  \BibitemOpen
  \bibfield  {author} {\bibinfo {author} {\bibfnamefont {A.~C.}\ \bibnamefont
  {Potter}}, \bibinfo {author} {\bibfnamefont {R.}~\bibnamefont {Vasseur}}, \
  and\ \bibinfo {author} {\bibfnamefont {S.~A.}\ \bibnamefont {Parameswaran}},\
  }\href {http://www.arxiv.org/abs/1501.03501} {\enquote {\bibinfo {title}
  {Universal properties of many-body delocalization transitions},}\ } (\bibinfo
  {year} {2015}),\ \Eprint {http://arxiv.org/abs/1501.03501} {1501.03501}
  \BibitemShut {NoStop}%
\bibitem [{\citenamefont {Luitz}\ \emph {et~al.}(2015)\citenamefont {Luitz},
  \citenamefont {Laflorencie},\ and\ \citenamefont {Alet}}]{Luitz2015}%
  \BibitemOpen
  \bibfield  {author} {\bibinfo {author} {\bibfnamefont {D.~J.}\ \bibnamefont
  {Luitz}}, \bibinfo {author} {\bibfnamefont {N.}~\bibnamefont {Laflorencie}},
  \ and\ \bibinfo {author} {\bibfnamefont {F.}~\bibnamefont {Alet}},\ }\href
  {\doibase 10.1103/PhysRevB.91.081103} {\bibfield  {journal} {\bibinfo
  {journal} {Phys. Rev. B}\ }\textbf {\bibinfo {volume} {91}},\ \bibinfo
  {pages} {081103} (\bibinfo {year} {2015})}\BibitemShut {NoStop}%
\bibitem [{\citenamefont {Berkelbach}\ and\ \citenamefont
  {Reichman}(2010)}]{Berkelbach2010}%
  \BibitemOpen
  \bibfield  {author} {\bibinfo {author} {\bibfnamefont {T.~C.}\ \bibnamefont
  {Berkelbach}}\ and\ \bibinfo {author} {\bibfnamefont {D.~R.}\ \bibnamefont
  {Reichman}},\ }\href {\doibase 10.1103/PhysRevB.81.224429} {\bibfield
  {journal} {\bibinfo  {journal} {Phys. Rev. B}\ }\textbf {\bibinfo {volume}
  {81}},\ \bibinfo {pages} {224429} (\bibinfo {year} {2010})}\BibitemShut
  {NoStop}%
\bibitem [{\citenamefont {Iyer}\ \emph {et~al.}(2013)\citenamefont {Iyer},
  \citenamefont {Oganesyan}, \citenamefont {Refael},\ and\ \citenamefont
  {Huse}}]{Iyer2013}%
  \BibitemOpen
  \bibfield  {author} {\bibinfo {author} {\bibfnamefont {S.}~\bibnamefont
  {Iyer}}, \bibinfo {author} {\bibfnamefont {V.}~\bibnamefont {Oganesyan}},
  \bibinfo {author} {\bibfnamefont {G.}~\bibnamefont {Refael}}, \ and\ \bibinfo
  {author} {\bibfnamefont {D.~A.}\ \bibnamefont {Huse}},\ }\href@noop {}
  {\bibfield  {journal} {\bibinfo  {journal} {Phys. Rev. B}\ }\textbf {\bibinfo
  {volume} {87}},\ \bibinfo {pages} {134202} (\bibinfo {year}
  {2013})}\BibitemShut {NoStop}%
\bibitem [{\citenamefont {Kjall}\ \emph {et~al.}(2014)\citenamefont {Kjall},
  \citenamefont {Bardarson},\ and\ \citenamefont {Pollmann}}]{Kjall2014}%
  \BibitemOpen
  \bibfield  {author} {\bibinfo {author} {\bibfnamefont {J.~A.}\ \bibnamefont
  {Kjall}}, \bibinfo {author} {\bibfnamefont {J.~H.}\ \bibnamefont
  {Bardarson}}, \ and\ \bibinfo {author} {\bibfnamefont {F.}~\bibnamefont
  {Pollmann}},\ }\href {\doibase 10.1103/PhysRevLett.113.107204} {\bibfield
  {journal} {\bibinfo  {journal} {Phys. Rev. Lett.}\ }\textbf {\bibinfo
  {volume} {113}},\ \bibinfo {pages} {107204} (\bibinfo {year}
  {2014})}\BibitemShut {NoStop}%
\bibitem [{\citenamefont {\ifmmode \check{Z}\else
  \v{Z}\fi{}nidari\ifmmode~\check{c}\else \v{c}\fi{}}\ \emph
  {et~al.}(2008)\citenamefont {\ifmmode \check{Z}\else
  \v{Z}\fi{}nidari\ifmmode~\check{c}\else \v{c}\fi{}}, \citenamefont {Prosen},\
  and\ \citenamefont {Prelov\ifmmode~\check{s}\else
  \v{s}\fi{}ek}}]{Znidaric2008}%
  \BibitemOpen
  \bibfield  {author} {\bibinfo {author} {\bibfnamefont {M.}~\bibnamefont
  {\ifmmode \check{Z}\else \v{Z}\fi{}nidari\ifmmode~\check{c}\else
  \v{c}\fi{}}}, \bibinfo {author} {\bibfnamefont {T.}~\bibnamefont {Prosen}}, \
  and\ \bibinfo {author} {\bibfnamefont {P.}~\bibnamefont
  {Prelov\ifmmode~\check{s}\else \v{s}\fi{}ek}},\ }\href {\doibase
  10.1103/PhysRevB.77.064426} {\bibfield  {journal} {\bibinfo  {journal} {Phys.
  Rev. B}\ }\textbf {\bibinfo {volume} {77}},\ \bibinfo {pages} {064426}
  (\bibinfo {year} {2008})}\BibitemShut {NoStop}%
\bibitem [{\citenamefont {Bardarson}\ \emph {et~al.}(2012)\citenamefont
  {Bardarson}, \citenamefont {Pollmann},\ and\ \citenamefont
  {Moore}}]{Bardarson2012}%
  \BibitemOpen
  \bibfield  {author} {\bibinfo {author} {\bibfnamefont {J.~H.}\ \bibnamefont
  {Bardarson}}, \bibinfo {author} {\bibfnamefont {F.}~\bibnamefont {Pollmann}},
  \ and\ \bibinfo {author} {\bibfnamefont {J.~E.}\ \bibnamefont {Moore}},\
  }\href {\doibase 10.1103/PhysRevLett.109.017202} {\bibfield  {journal}
  {\bibinfo  {journal} {Phys. Rev. Lett.}\ }\textbf {\bibinfo {volume} {109}},\
  \bibinfo {pages} {017202} (\bibinfo {year} {2012})}\BibitemShut {NoStop}%
\bibitem [{\citenamefont {Serbyn}\ \emph
  {et~al.}(2013{\natexlab{b}})\citenamefont {Serbyn}, \citenamefont
  {Papi\ifmmode~\acute{c}\else \'{c}\fi{}},\ and\ \citenamefont
  {Abanin}}]{Serbyn2013a}%
  \BibitemOpen
  \bibfield  {author} {\bibinfo {author} {\bibfnamefont {M.}~\bibnamefont
  {Serbyn}}, \bibinfo {author} {\bibfnamefont {Z.}~\bibnamefont
  {Papi\ifmmode~\acute{c}\else \'{c}\fi{}}}, \ and\ \bibinfo {author}
  {\bibfnamefont {D.~A.}\ \bibnamefont {Abanin}},\ }\href {\doibase
  10.1103/PhysRevLett.110.260601} {\bibfield  {journal} {\bibinfo  {journal}
  {Phys. Rev. Lett.}\ }\textbf {\bibinfo {volume} {110}},\ \bibinfo {pages}
  {260601} (\bibinfo {year} {2013}{\natexlab{b}})}\BibitemShut {NoStop}%
\bibitem [{\citenamefont {Chandran}\ \emph {et~al.}(2014)\citenamefont
  {Chandran}, \citenamefont {Carrasquilla}, \citenamefont {Abanin},\ and\
  \citenamefont {Vidal}}]{Chandran2014}%
  \BibitemOpen
  \bibfield  {author} {\bibinfo {author} {\bibfnamefont {A.}~\bibnamefont
  {Chandran}}, \bibinfo {author} {\bibfnamefont {I.}~\bibnamefont
  {Carrasquilla}, \bibfnamefont {Kim}}, \bibinfo {author} {\bibfnamefont
  {D.}~\bibnamefont {Abanin}}, \ and\ \bibinfo {author} {\bibfnamefont
  {G.}~\bibnamefont {Vidal}},\ }\href@noop {} {\bibfield  {journal} {\bibinfo
  {journal} {arXiv:1410.0687}\ } (\bibinfo {year} {2014})}\BibitemShut
  {NoStop}%
\bibitem [{\citenamefont {Pekker}\ and\ \citenamefont
  {Clark}(2014)}]{Pekker2014a}%
  \BibitemOpen
  \bibfield  {author} {\bibinfo {author} {\bibfnamefont {D.}~\bibnamefont
  {Pekker}}\ and\ \bibinfo {author} {\bibfnamefont {B.~K.}\ \bibnamefont
  {Clark}},\ }\href {http://www.arxiv.org/abs/1410.2224} {\enquote {\bibinfo
  {title} {Encoding the structure of many-body localization with matrix product
  operators},}\ } (\bibinfo {year} {2014}),\ \Eprint
  {http://arxiv.org/abs/1410.2224} {1410.2224} \BibitemShut {NoStop}%
\bibitem [{\citenamefont {Pollmann}\ \emph {et~al.}(2015)\citenamefont
  {Pollmann}, \citenamefont {Khemani}, \citenamefont {Cirac},\ and\
  \citenamefont {Sondhi}}]{Pollmann2015}%
  \BibitemOpen
  \bibfield  {author} {\bibinfo {author} {\bibfnamefont {F.}~\bibnamefont
  {Pollmann}}, \bibinfo {author} {\bibfnamefont {V.}~\bibnamefont {Khemani}},
  \bibinfo {author} {\bibfnamefont {J.~I.}\ \bibnamefont {Cirac}}, \ and\
  \bibinfo {author} {\bibfnamefont {S.~L.}\ \bibnamefont {Sondhi}},\ }\href
  {http://www.arxiv.org/abs/1506.07179} {\enquote {\bibinfo {title} {Efficient
  variational diagonalization of fully many-body localized hamiltonians},}\ }
  (\bibinfo {year} {2015}),\ \Eprint {http://arxiv.org/abs/1506.07179}
  {1506.07179} \BibitemShut {NoStop}%
\bibitem [{\citenamefont {D'Errico}\ \emph {et~al.}(2014)\citenamefont
  {D'Errico}, \citenamefont {Lucioni}, \citenamefont {Tanzi}, \citenamefont
  {Gori}, \citenamefont {Roux}, \citenamefont {McCulloch}, \citenamefont
  {Giamarchi}, \citenamefont {Inguscio},\ and\ \citenamefont
  {Modugno}}]{DErrico2014}%
  \BibitemOpen
  \bibfield  {author} {\bibinfo {author} {\bibfnamefont {C.}~\bibnamefont
  {D'Errico}}, \bibinfo {author} {\bibfnamefont {E.}~\bibnamefont {Lucioni}},
  \bibinfo {author} {\bibfnamefont {L.}~\bibnamefont {Tanzi}}, \bibinfo
  {author} {\bibfnamefont {L.}~\bibnamefont {Gori}}, \bibinfo {author}
  {\bibfnamefont {G.}~\bibnamefont {Roux}}, \bibinfo {author} {\bibfnamefont
  {I.~P.}\ \bibnamefont {McCulloch}}, \bibinfo {author} {\bibfnamefont
  {T.}~\bibnamefont {Giamarchi}}, \bibinfo {author} {\bibfnamefont
  {M.}~\bibnamefont {Inguscio}}, \ and\ \bibinfo {author} {\bibfnamefont
  {G.}~\bibnamefont {Modugno}},\ }\href {\doibase
  10.1103/PhysRevLett.113.095301} {\bibfield  {journal} {\bibinfo  {journal}
  {Phys. Rev. Lett.}\ }\textbf {\bibinfo {volume} {113}},\ \bibinfo {pages}
  {095301} (\bibinfo {year} {2014})}\BibitemShut {NoStop}%
\bibitem [{\citenamefont {Meldgin}\ \emph {et~al.}(2015)\citenamefont
  {Meldgin}, \citenamefont {Ray}, \citenamefont {Russ}, \citenamefont
  {Ceperley},\ and\ \citenamefont {DeMarco}}]{Meldgin2015}%
  \BibitemOpen
  \bibfield  {author} {\bibinfo {author} {\bibfnamefont {C.}~\bibnamefont
  {Meldgin}}, \bibinfo {author} {\bibfnamefont {U.}~\bibnamefont {Ray}},
  \bibinfo {author} {\bibfnamefont {P.}~\bibnamefont {Russ}}, \bibinfo {author}
  {\bibfnamefont {D.}~\bibnamefont {Ceperley}}, \ and\ \bibinfo {author}
  {\bibfnamefont {B.}~\bibnamefont {DeMarco}},\ }\href
  {http://www.arxiv.org/abs/1502.02333} {\enquote {\bibinfo {title} {Probing
  the bose-glass--superfluid transition using quantum quenches of disorder},}\
  } (\bibinfo {year} {2015}),\ \Eprint {http://arxiv.org/abs/1502.02333}
  {1502.02333} \BibitemShut {NoStop}%
\bibitem [{\citenamefont {Schreiber}\ \emph {et~al.}(2015)\citenamefont
  {Schreiber}, \citenamefont {Hodgman}, \citenamefont {Bordia}, \citenamefont
  {L{\"u}schen}, \citenamefont {Fischer}, \citenamefont {Vosk}, \citenamefont
  {Altman}, \citenamefont {Schneider},\ and\ \citenamefont
  {Bloch}}]{Schreiber2015}%
  \BibitemOpen
  \bibfield  {author} {\bibinfo {author} {\bibfnamefont {M.}~\bibnamefont
  {Schreiber}}, \bibinfo {author} {\bibfnamefont {S.~S.}\ \bibnamefont
  {Hodgman}}, \bibinfo {author} {\bibfnamefont {P.}~\bibnamefont {Bordia}},
  \bibinfo {author} {\bibfnamefont {H.~P.}\ \bibnamefont {L{\"u}schen}},
  \bibinfo {author} {\bibfnamefont {M.~H.}\ \bibnamefont {Fischer}}, \bibinfo
  {author} {\bibfnamefont {R.}~\bibnamefont {Vosk}}, \bibinfo {author}
  {\bibfnamefont {E.}~\bibnamefont {Altman}}, \bibinfo {author} {\bibfnamefont
  {U.}~\bibnamefont {Schneider}}, \ and\ \bibinfo {author} {\bibfnamefont
  {I.}~\bibnamefont {Bloch}},\ }\href {\doibase 10.1126/science.aaa7432}
  {\bibfield  {journal} {\bibinfo  {journal} {Science}\ }\textbf {\bibinfo
  {volume} {349}},\ \bibinfo {pages} {842} (\bibinfo {year} {2015})},\ \Eprint
  {http://arxiv.org/abs/http://www.sciencemag.org/content/349/6250/842.full.pdf}
  {http://www.sciencemag.org/content/349/6250/842.full.pdf} \BibitemShut
  {NoStop}%
\bibitem [{\citenamefont {Swingle}(2013)}]{Swingle2013}%
  \BibitemOpen
  \bibfield  {author} {\bibinfo {author} {\bibfnamefont {B.}~\bibnamefont
  {Swingle}},\ }\href@noop {} {\bibfield  {journal} {\bibinfo  {journal}
  {arXiv:1307.0507}\ } (\bibinfo {year} {2013})}\BibitemShut {NoStop}%
\bibitem [{\citenamefont {Friesdorf}\ \emph {et~al.}(2014)\citenamefont
  {Friesdorf}, \citenamefont {Werner}, \citenamefont {Brown}, \citenamefont
  {Scholz},\ and\ \citenamefont {Eisert}}]{Friesdorf2014}%
  \BibitemOpen
  \bibfield  {author} {\bibinfo {author} {\bibfnamefont {M.}~\bibnamefont
  {Friesdorf}}, \bibinfo {author} {\bibfnamefont {A.~H.}\ \bibnamefont
  {Werner}}, \bibinfo {author} {\bibfnamefont {W.}~\bibnamefont {Brown}},
  \bibinfo {author} {\bibfnamefont {V.~B.}\ \bibnamefont {Scholz}}, \ and\
  \bibinfo {author} {\bibfnamefont {J.}~\bibnamefont {Eisert}},\ }\href@noop {}
  {\bibfield  {journal} {\bibinfo  {journal} {arXiv:1409.1252}\ } (\bibinfo
  {year} {2014})}\BibitemShut {NoStop}%
\bibitem [{\citenamefont {Schollw{\"o}ck}(2011)}]{Schollwock2011}%
  \BibitemOpen
  \bibfield  {author} {\bibinfo {author} {\bibfnamefont {U.}~\bibnamefont
  {Schollw{\"o}ck}},\ }\href@noop {} {\bibfield  {journal} {\bibinfo  {journal}
  {Annals of Physics}\ }\textbf {\bibinfo {volume} {326}},\ \bibinfo {pages}
  {96} (\bibinfo {year} {2011})}\BibitemShut {NoStop}%
\bibitem [{\citenamefont {Deutsch}(1991)}]{Deutsch1991}%
  \BibitemOpen
  \bibfield  {author} {\bibinfo {author} {\bibfnamefont {J.~M.}\ \bibnamefont
  {Deutsch}},\ }\href {\doibase 10.1103/PhysRevA.43.2046} {\bibfield  {journal}
  {\bibinfo  {journal} {Phys. Rev. A}\ }\textbf {\bibinfo {volume} {43}},\
  \bibinfo {pages} {2046} (\bibinfo {year} {1991})}\BibitemShut {NoStop}%
\bibitem [{\citenamefont {Srednicki}(1994)}]{Srednicki1994}%
  \BibitemOpen
  \bibfield  {author} {\bibinfo {author} {\bibfnamefont {M.}~\bibnamefont
  {Srednicki}},\ }\href {\doibase 10.1103/PhysRevE.50.888} {\bibfield
  {journal} {\bibinfo  {journal} {Phys. Rev. E}\ }\textbf {\bibinfo {volume}
  {50}},\ \bibinfo {pages} {888} (\bibinfo {year} {1994})}\BibitemShut
  {NoStop}%
\bibitem [{\citenamefont {Rigol}\ \emph {et~al.}(2008)\citenamefont {Rigol},
  \citenamefont {Dunjko},\ and\ \citenamefont {Olshanii}}]{Rigol2008}%
  \BibitemOpen
  \bibfield  {author} {\bibinfo {author} {\bibfnamefont {M.}~\bibnamefont
  {Rigol}}, \bibinfo {author} {\bibfnamefont {V.}~\bibnamefont {Dunjko}}, \
  and\ \bibinfo {author} {\bibfnamefont {M.}~\bibnamefont {Olshanii}},\ }\href
  {http://dx.doi.org/10.1038/nature06838} {\bibfield  {journal} {\bibinfo
  {journal} {Nature}\ }\textbf {\bibinfo {volume} {452}},\ \bibinfo {pages}
  {854} (\bibinfo {year} {2008})}\BibitemShut {NoStop}%
\bibitem [{\citenamefont {Grover}(2014)}]{Grover2014}%
  \BibitemOpen
  \bibfield  {author} {\bibinfo {author} {\bibfnamefont {T.}~\bibnamefont
  {Grover}},\ }\href {http://www.arxiv.org/abs/1405.1471} {\enquote {\bibinfo
  {title} {Certain general constraints on the many-body localization
  transition},}\ } (\bibinfo {year} {2014}),\ \Eprint
  {http://arxiv.org/abs/1405.1471} {1405.1471} \BibitemShut {NoStop}%
\bibitem [{\citenamefont {Khemani}\ \emph {et~al.}(2015)\citenamefont
  {Khemani}, \citenamefont {Pollmann},\ and\ \citenamefont
  {Sondhi}}]{Khemani2015}%
  \BibitemOpen
  \bibfield  {author} {\bibinfo {author} {\bibfnamefont {V.}~\bibnamefont
  {Khemani}}, \bibinfo {author} {\bibfnamefont {F.}~\bibnamefont {Pollmann}}, \
  and\ \bibinfo {author} {\bibfnamefont {S.~L.}\ \bibnamefont {Sondhi}},\
  }\href {http://www.arxiv.org/abs/1509.00483} {\enquote {\bibinfo {title}
  {Obtaining highly-excited eigenstates of many-body localized hamiltonians by
  the density matrix renormalization group},}\ } (\bibinfo {year} {2015}),\
  \Eprint {http://arxiv.org/abs/1509.00483} {1509.00483} \BibitemShut {NoStop}%
\bibitem [{\citenamefont {Hu}\ and\ \citenamefont
  {Chan}(2015)}]{hu2015excited}%
  \BibitemOpen
  \bibfield  {author} {\bibinfo {author} {\bibfnamefont {W.}~\bibnamefont
  {Hu}}\ and\ \bibinfo {author} {\bibfnamefont {G.~K.}\ \bibnamefont {Chan}},\
  }\href@noop {} {\bibfield  {journal} {\bibinfo  {journal} {Journal of
  Chemical Theory and Computation}\ } (\bibinfo {year} {2015})}\BibitemShut
  {NoStop}%
\end{thebibliography}%
\end{document}